# Terahertz Magnetic Modulator based on Magnetically-Clustered Nanoparticles

Mostafa Shalaby,[1,2*] Marco Peccianti,[3,] Yavuz Ozturk[1], Ibraheem Al-Naib,[1] Christoph P. Hauri,[2,4] and Roberto Morandotti[1*]

[1]*INRS-EMT, Varennes, Quebec J3X 1S2, Canada*

[2]*SwissFEL, Paul Scherrer Institute, 5232 Villigen PSI, Switzerland*

[3]*Department of Physics and Astronomy, University of Sussex, Pevensey Building II, 3A8, Falmer, Brighton BN1 9QH, United Kingdom*

[4]*Ecole Polytechnique Federale de Lausanne, 1015 Lausanne, Switzerland*

(*most.shalaby@gmail.com and morandotti@emt.inrs.ca)

*Random orientation of liquid-suspended magnetic nanoparticles (Ferrofluid) gives rise to zero net magnetic orientation. An external magnetic field tends to align them into clusters, leading to a strong linear dichroism on a propagating wave. Using 10 nm-sized $Fe_3O_4$, we experimentally realize a polarization-sensitive magnetic modulator operating at terahertz wavelengths. We reached a modulation depth of 66% using a field of 35 mT. The proposed concept offers a solution towards fundamental terahertz magnetic modulators.*

Terahertz (THz) signal processing recently rose to prominence with countless potential applications. On the one side, the increasing demand for high bandwidth and data rate in sub-THz wireless communication systems keeps pushing up the frequency limit, reaching the edge of the THz band[1]. On the other hand, great efforts are being made to extend the well-established infrared materials spectroscopic techniques to the THz regime. Terahertz is capable of exceptional matter composition discrimination due to the inherently compound-dependent fingerprints exhibited in this bandwidth[2-8].

Although sources[9-13] and detectors technologies[14] have evolved rapidly over the past years, THz radiation is still difficult to manipulate mainly because of the lack of both suitable materials and efficient modulation (control) techniques. Terahertz modulation has been demonstrated by optical[15-20], electronic[21-23], and thermal[24-25] means. Those techniques differ in bandwidth, complexity, flexibility and modulation depth, hence preferences are usually dictated by application constraints. For instance, optical beams can dramatically change the electric current density in a semiconductor and modulate a propagating THz pulse in fractions of a picosecond, but this technique depends on the availability of intense ultrashort femtosecond laser sources. On the other hand, THz modulation through a $VO_2$ film can be triggered just heating up the sample with a small electric current flowing in a conducting wire. However, material response is limited to the scale of tens of a millisecond.

Magnetic fields are an important tool to change the material response against a propagating electromagnetic wave, but their effect is generally weak[26]. Efficient modulation thus requires a significant propagation length, which is prevented by the associated losses (which generally increase with frequency). Hence, a practical THz magnetic modulator has not been realized so far.

In this paper, we use liquid-suspended magnetic nanoparticles (i.e., a Ferrofluid[27]) to achieve an efficient modulation of THz pulses using very low magnetic fields. A modulation depth as high as 66% is shown using a magnetic field as low as 35 mT. The new concept proposed here may open an alternative and perhaps a new paradigm for future THz modulation devices and systems.

The sample consists of a Ferrofluid-filled 10 *mm*-long cuvette. Our Ferrofluid is commercially provided and consists of 10 *nm*-sized $Fe_3O_4$ particles suspended in a carrier liquid. The nanoparticles are coated with a stabilizing surfactant providing electrostatic resistance against agglomeration and thus preserving their free movement as non-interacting particles. This property makes them sensitive to small (milli-Tesla level) magnetic fields. A non-uniform (spatially-varying) magnetic field can impose a strong force on those nanoparticles that not only rotates them but can also sweep them along the field gradient. On the contrary, uniform fields tend to simply align the nanoparticles along the field direction. Typically, two kinds of contribution to the magnetic moment reorientation can be recognized here: Brownian and Neel. While the former tends to physically rotate the particles to the field direction, the latter just rotates the magnetic moments without any physical rotation [28]. The otherwise randomly-oriented (Fig. 1a) particles appear to be organized in the form of clusters in the direction of the field lines (Figs. 1b and 1c).

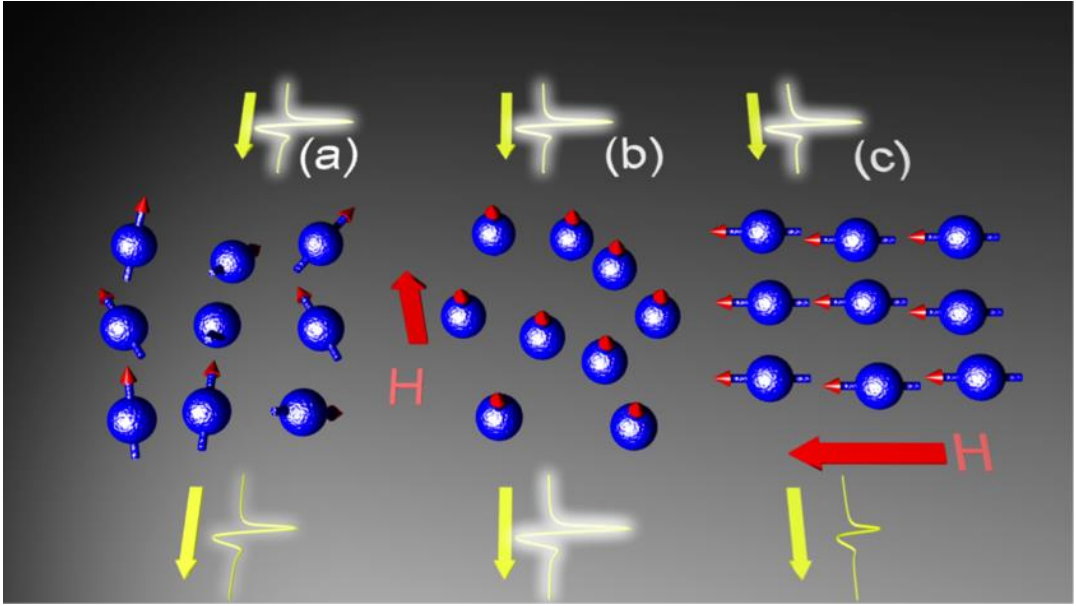

**Figure 1.** Nanoparticles alignment with the external static magnetic field (H) and its effect on THz propagation. (a) In the absence of an external field, the nanoparticles are randomly oriented giving rise to a zero magnetic state and the THz experiences isotropic absorption. (b) & (c) An external magnetic field tends to align the particles along its direction inducing THz linear dichroism. If the particles orientation is orthogonal/parallel to the THz electric field (b)/(c), a lower/higher absorption is expected.

Cluster formation is a basic mechanism responsible for many of the unique properties Ferrofluids exhibit[29]. For example, if the particles get aligned along the direction of wave propagation, this builds up net magnetization (M) in the same direction, which in turn leads to a difference in the propagation velocity of the wave circular eigenmodes and thus in the rotation of the plane of polarization (Faraday rotation)[30-31]. An in-plane magnetic field induces a directional absorption (linear dichroism) of a propagating electromagnetic wave. This interesting phenomenon has interesting consequences. For example, following our measurements of the in-plane magnetic properties of Ferrofluids at THz frequencies,[32] Chen *et al*, measured the H-induced tunability of the in-plane real refractive index[33]. Here, we magnetically control the THz absorption to modulate broadband THz pulses.

Magnetic particle alignment and cluster formation induce a variation in the absorption coefficient $\Delta\alpha = \alpha(H) - \alpha(0)$, strongly dependent on the angle between the cluster axis (external magnetic field direction) and the THz electric polarization. $\alpha(H)$ and $\alpha(0)$ are the absorption coefficients in the presence and absence of an external magnetic field H, respectively. Two main absorption mechanisms can be responsible for the attenuation[34]: (a) Absorption by the propagating field induced imaginary magnetic polarizations (Eddy currents losses)- which is ignored here because of the low macroscopic conductivity between magnetic nanoparticles, (b) Absorption by the field induced imaginary electric polarization. This latter component represents the current generated within the colloidal nanoparticles. Even in the presence of weak magnetic fields, this component can lead to significant attenuation and is thus the main mechanism responsible for the light absorption considered here. A propagating wave with the electric field polarized parallel to the cluster orientation - *the extraordinary wave*, undergoes absorption (Fig. 1c), as opposite to the non-clustered (randomly oriented particles) case (Fig. 1a). At the same time, a wave polarized orthogonal to the cluster direction (Fig. 1b) - *the ordinary wave*- undergoes a reduced attenuation and shows an increase in transmission relative to the reference (isotropic) case (Fig. 1a).

We performed our experimental measurements using a time domain terahertz spectroscopy setup. The laser pulses (energy = ~2 mJ, duration = 130 fs, repetition rate = 1 kHz, center wavelength = 800 nm) were split between the terahertz generation -through optical rectification-and detection -via electro-optical sampling- in two different ZnTe crystals. The sample is placed in the x-y plane and z is taken to be the direction of propagation. We used the EFH ferrofluid series (EFH1 and EFH3 with particles concentrations of 7.8% and 12.4%, respectively) because of their organic solvent that exhibits significantly lower absorption in the THz band when compared to water-based Ferrofluids. Unless otherwise stated, EFH1 was used.

To demonstrate the magnetic field-induced dichroism, we consistently probed the transmission of the ordinary and extraordinary wave components. We placed the sample in an x-aligned (planar) magnetic field generated using an electromagnet (GMW-3470). To measure the transmitted extraordinary wave, the THz was (horizontally) x-polarized. Two (y-) vertically-aligned wire grid polarizers were placed before and after the sample to ensure the THz horizontal polarization. In the case of an ordinary wave measurement, THz generation, detection and wire grid polarizers were rotated by 90°. Figure 2 shows the transmitted pulses and the corresponding spectra of the extraordinary and ordinary waves for several magnetic field levels, specifically 0, 8, 17, 35, and 106 mT.

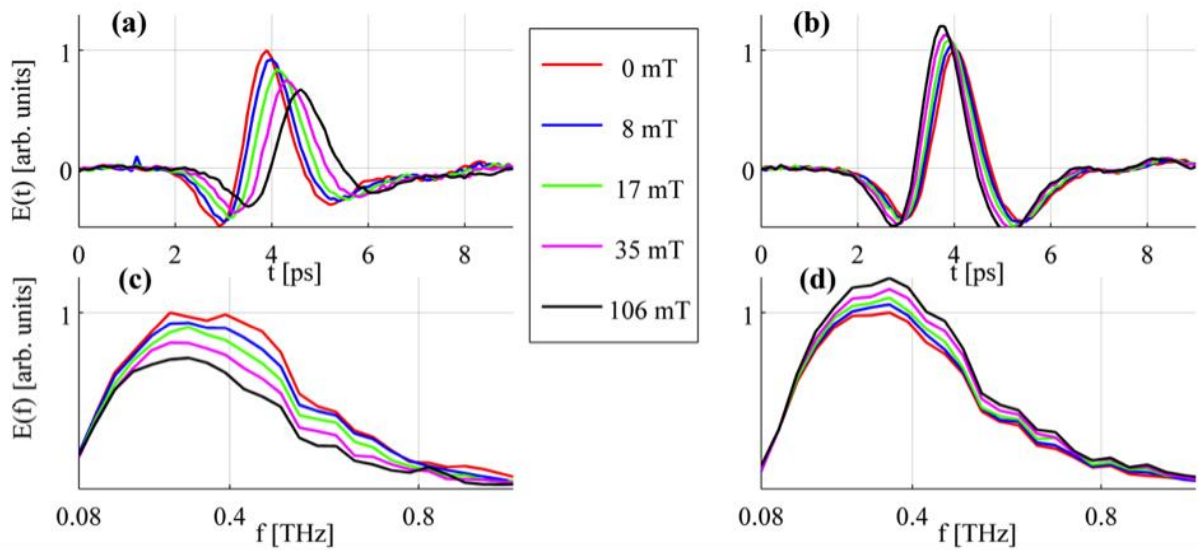

**Figure 2.** Transmitted THz waves under the application of different external magnetic fields. When the THz is polarized parallel to the applied field (a) & (c), a strong attenuation is observed. A THz polarized orthogonal to the external magnetic field shows an increase in transmission (b) & (d) in comparison with the zero-field randomly oriented case. The rate of induced attenuation decreases with the increase in the applied magnetic field.

With the increase in the magnetic field, the transmitted extraordinary/ordinary polarization decreases/increases confirming the dichroism over the broad THz spectrum. This is accompanied by a magnetic field-induced birefringence. The magnetic field-induced absorption coefficients of the extraordinary ($\parallel$) and ordinary ($\perp$) polarizations are related by

$$\Delta\alpha_{\parallel} = -2\Delta\alpha_{\perp} \qquad (1)$$

This relation was verified using near infrared probing[29]. However, it is purely related to the average domain reorientation and does not depend on the frequency as long as the wavelength is greater than the *nm* scale of the particle chain. The measured change in transmission can originate from both the change in the Fresnel reflection losses at the interfaces and the bulk attenuation. The perturbation in the complex refractive index of the sample was found to be less than 3% for the levels of the magnetic fields in our experiment. We, therefore, assume no change in the Fresnel losses and the bulk losses are solely responsible for the change in the transmission. To extract the magnetic field-induced absorption ($\Delta\alpha_{\parallel}$ and $\Delta\alpha_{\perp}$) from the experimental measurement, we first write the spectral components of the THz field as $E_t(\omega) = E_0(\omega)e^{-\alpha d}$ where $d = 10\ mm$ is the sample thickness. $E_t$ and $E_0$ are the modulated and unmodulated fields, respectively. From this, $\Delta\alpha$ can be readily extracted using the logarithmic transmission $t_1 = \ln\frac{E_t(\omega)_i}{E_0(\omega)} = -\Delta\alpha_i d$, $i \in (\parallel, \perp)$. Figure 3a shows the extracted extraordinary-induced absorption for two levels of magnetization (17 mT and 35 mT). The good agreement over the wide THz bandwidth with the prediction obtained from Eq. 1 is also shown.

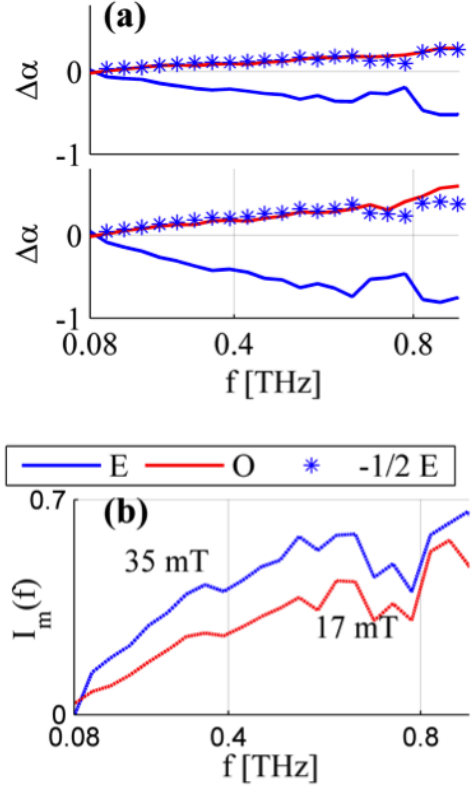

**Figure 3.** The induced absorption of the extraordinary (E) and ordinary (O) waves for applied fields of 17 mT (top) and 35 mT (bottom). The experimental measurements are shown in blue and red solid lines. Asterisks show the E-wave measurement after applying Eq. 1 to calculate the attenuation in the O-wave. (b) Modulation depth of the E-wave calculated using Eq. 2 for the two magnetic field levels.

To evaluate the efficiency of the modulation process we calculate the energy spectral density modulation depth

$$I_m(\omega) = \frac{|E_0(\omega)|^2 - |E_t(\omega)|^2}{|E_0(\omega)|^2} \qquad (2)$$

The frequency-resolved modulation intensity is presented in Fig. 3b for the extraordinary wave at the two levels of the magnetic field (17 mT and 35 mT) shown in Fig. 3a where up to 66% modulation is found for a field of 35 mT. We stress here that in a perspective modulation device, the required (very) low magnetic field can be locally obtained by a moderate current flowing in a wire.

The modulation increases with both frequency and applied magnetic field. In principle, the induced magnetization and thus modulation should continue to increase until saturation (~1T for EFH1). However, at higher magnetic field levels, the magnetization build up has a nonlinear trend (saturates). This behavior is described by the Langevin relation $M = coth(KH) - 1/KH$ where K is a temperature-dependent parameter[18]. As $\Delta\alpha_\parallel$ is proportional to M, the indued absorption is expected to have a similar Langevian dependence. This is experimentally demonstrated in Fig. 4a where both M and the extraordinary $\Delta\alpha_\parallel$ are shown for fields up to 600 mT. We would like to emphasize here that EFH1 requires 1T to reach the magnetization saturation ($M_s = 40$ mT). Yet, due to the nonlinear behavior, only 30 mT is required to reach $M_s/2$ (with a magnetization approximately linear with the applied field up to this level).

The attenuation process is mediated by an increase in the electrical conductivity with the increase of the number of particles aligned with the magnetic field. The modulation process is thus expected to be independent of the THz polarity. This is confirmed in Fig. 4b where the THz pulses are measured under two equal but oppositely-polarized magnetic fields. Finally, the effect of the nanoparticles concentration is briefly considered here by comparing two Ferrofluids from the same series, EFH3 and EFH1. The particle concentration of the former is 1.5 times higher than that of the latter.

The induced absorption is directly proportional to the particle concentration[29] and so is $\Delta\alpha_{EFH3} = 1.5 \times \Delta\alpha_{EFH1}$. This last relation is experimentally verified and shown in Fig. 4c, which demonstrates an excellent agreement with the theoretical predictions. This implies that a higher absorption modulation can be obtained by increasing the concretion of the sample. However, this comes at the expense of higher absorption. The advantage of using higher concentration liquids over longer sample lengths can be seen if this liquid is coupled with other structures where the thickness can not be arbitrarily varied (like metamaterials) or where a higher thickness induces more losses associated with the structure itself (like waveguides).

In conclusion, we demonstrated terahertz magnetic modulation using magnetic field-induced clustering of nanoparticles in Ferrofluids. The demonstrated technique combines a high modulation depth and low magnetic field requirements while preserving the flexibility given by the possibility of using liquids. We believe that our results will pave the way to a new class of THz modulators that can be integrated in other magnetic/nonmagnetic systems such as metamaterials and waveguides.

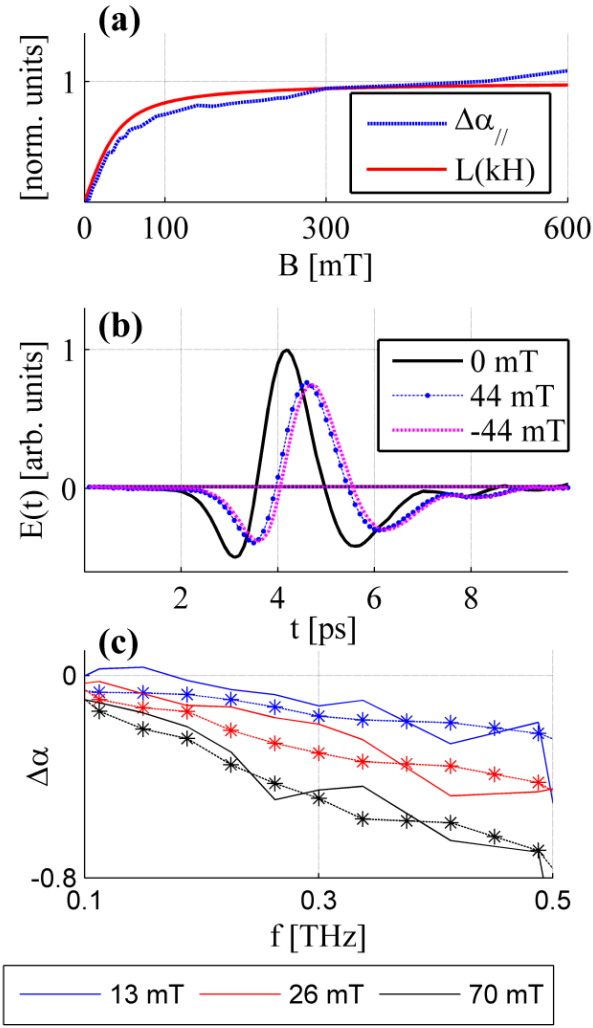

**Figure 4.** Langevian behaviour of both the magnetization and the induced THz absorption of the E-wave. (b) Waveforms of the THz E-wave in the absence of an external field and in the case of two equal but oppositely-polarized fields. (c) The induced absorption of the E-wave in EFH3 (solid lines) and EFH1 after scaling by 1.5 (asterisks) to account for the difference in concentration. The agreement between the plot pairs demonstrates the scalability of the induced absorption by the concentration.

References

[1] T. Kleine-Ostmann and T. Nagatsuma, *J. Infrared Milli. Terahz. Waves* **32**, 143, 2011.

[2] W. L. J. Deibel and D. M. Mittleman, *Rep. Prog. Phys.*. **70**, 1325, 2007.

[3] B. B. Hu and M. C. Nuss, *Opt. Lett.* **20**, 1716, 1995.

[4] X. –C. Zhang, *Phys. Med. Biol.* **47**, 3667, 2002.

[5] M. R. Leahy-Hoppa, M. J. Fitch, X. Zheng, and L. M. Hayden, *Chem. Phys. Lett*. **434**, 227, 2007.

[6] K. Tanaka, H. Hirori, and M. Nagai, *IEEE Trans. Terahertz Sci. Technol.* **1**, 301, 2011.

[7] T. Kampfrath, *et al*., *Nat. Photonics* **5**, 31, 2011.

[8] M. Liu, *et al, Nature* **487**, 345, 2012.

[9] M. Shalaby and C. P. Hauri, arXiv: 1407.1656

[10] D. Daranciang, *et al*., *Appl. Phys. Lett.*. **99**, 141117, 2011.

[11] H. Hirori, A. Doi, F. Blanchard, and K. Tanaka, *Appl. Phys. Lett.* **98**, 091106, 2011.

[12] C.P. Hauri, C. Ruchert, C. Vicario, and F. Ardana, *Appl. Phys. Lett*. **99**, 161116 (2011).

[13] W. Shi, Y. Ding, N. Fernelius, and K. Vodopyanov, *Opt. Lett.* **27**, 1454, 2002.

[14] J. Dai, X. Xie and X. –C. Zhang, *Phys. Rev. Lett.* **97**, 103903, 2006.

[15] C. Janke, J. G. Rivas, P. H. Bolivar, and H. Kurz, *Opt. Lett.* **30**, 2357, 2005

[16] E. Hendry, M. J. Lockyear, J. G. Rivas, L. Kuipers, and M. Bonn, *Phys. Rev. B* **75**, 235305, 2007.

[17] W. J. Padilla, A. J. Taylor, C. Highstrete, M. Lee, and R. D. Averitt, *Phys. Rev. Lett.* **96**, 107401, 2006.

[18] D. R. Chowdhury, *et al.*, *Appl. Phys. Lett.* **99**, 231101, 2011.

[19] A. E. Nikolaenko, *et al., Opt. Express* **20**, 6068, 2012.

[20] T. Kleine-Ostmann, P. Dawson, K. Pierz, G. Hein, and M. Koch, *Appl. Phys. Lett.* **84**, 3555, 2004.

[21] H. −T. Chen, *et al., Nature* **444**, 597, 2006.

[22] W. L. Chan, *et al.*, *Appl. Phys. Lett.* **94**, 213511, 2009.

[23] H. −T. Chen, *et al., Nat. Photonics* **3**, 148, 2009.

[24] T. Driscoll, *et al., Appl. Phys. Lett.* **93**, 024101, 2008.

[25] R. Singh, *et al., Opt. Lett.* **36**, 1230, 2011.

[26] M. Shalaby, *et al.,* Phys. Rev. B **88**, 140301(R), 2013.

[27] S. S. Papell, *U.S. Patent* 3,215,572, 1965.

[28] W. F. Brown, *J. Appl. Phys.* **34**, 1319, 1963; Neel, L. *Ann. Geophys.* **5**, 99, 1949.

[29] N. N. Inaba, H. Miyajima, H. Takahashi, S. Taketomi, and S. Chikazumi, *IEEE Trans. on Magn.* **25**, 3866, 1989.

[30] M. Shalaby, *et al., Appl. Phys. Lett.* **100**, 241107, 2012.

[31] M. Shalaby, M. Peccianti, Y. Ozturk, and R. Morandotti, Nat. Commun. **4**, 1552, 2013.

[32] M. Shalaby, *et al.*, "Polarization-sensitive Magnetic Field Induced Modulation of Broadband THz Pulses in Liquid," CLEO: Science and Innovations San Jose, California, United States, 2012.

[33] S. Chen et al., *Opt. Express* **22**, 6313-6321 (2014).

[34] S. Taketomi, S. Ogawa, H. Miyajima, and S. Chikazumi, S. *IEEE Translation J. on Magn. in Japan* **4**, 384, 1989.